\documentclass[epj]{svjour}
\usepackage{graphics}
\begin{document}
\title{Hyperon-mixed neutron star with universal many-body repulsion}
\author{Y. Yamamoto\inst{1}, T. Furumoto\inst{2}, N. Yasutake\inst{3}
\and Th.A. Rijken\inst{4}% etc
}                     % Do not remove
\offprints{}          % Insert a name or remove this line
\institute{Nishina Center for Accelerator-Based Science,
Institute for Physical and Chemical
Research (RIKEN), Wako, Saitama, 351-0198, Japan
\and National Institute of Technology, Ichinoseki College, Ichinoseki, 
Iwate, 021-8511, Japan
\and Department of Physics, Chiba Institute of Technology, 2-1-1 Shibazono
Narashino, Chiba 275-0023, Japan
\and IMAPP, University of Nijmegen, Nijmegen, The Netherlands}
\date{Received: date / Revised version: date}
\abstract{Neutron stars with large masses $\sim 2M_{\odot}$
require the hard stiffness of equation of state (EoS) of 
neutron-star matter. On the other hand, hyperon mixing 
brings about remarkable softening of EoS.
In order to solve this problem, a multi-pomeron exchange potential 
(MPP) is introduced as a model for the universal many-body repulsion 
in baryonic systems on the basis of the Extended Soft Core (ESC) 
baryon-baryon interaction. 
The strength of MPP is determined by analyzing the nucleus-nucleus 
scattering with the G-matrix folding model. 
The interactions in $\Lambda\!N$, $\Sigma\!N$ and $\Xi\!N$ channels 
are shown to be consistent with experimental indications.
The EoS in neutron-star matter with hyperon mixing is obtained 
from ESC in addition of MPP, 
and mass-radius relations of neutron stars are derived.
The maximum mass is shown to reach $2M_{\odot}$ even in the case of 
including hyperon mixing on the basis of model-parameters determined 
by terrestrial experiments.
\PACS{
{21.65.Cd} {21.80.+a} {25.70.-z} {26.60.Kp}
%      \and
%      {PACS-key}{discribing text of that key}
     } % end of PACS codes
} %end of abstract
\maketitle
\section{Introduction}
\label{intro}
The observed masses of neutron stars J1614-2230~\cite{Demorest10} and
J0348-0432~\cite{Antoniadis13} are given as $(1.97\pm0.04)M_{\odot}$ 
and $(2.01\pm0.04)M_{\odot}$, respectively.
These large masses give a severe condition for the stiffness of
equation of state (EoS) of neutron-star matter.
It is well known that the stiff EoS giving the maximum mass of 
$2M_{\odot}$ can be derived from the existence of strong 
three-nucleon repulsion (TNR) in the high-density region.
However, the hyperon ($Y$) mixing in neutron-star matter 
brings about the remarkable softening of the EoS, which 
cancels the TNR effect for the maximum mass~\cite{Baldo00,Vidana00,NYT}.

One of ideas to avoid this serious problem, called "Hyperon puzzle 
in neutron stars", is to consider that the TNR-like repulsions work 
universally for $Y\!N\!N$, $Y\!Y\!N$ $Y\!Y\!Y$ as well as for 
$N\!N\!N$ \cite{NYT}.
In our previous works~\cite{YFYR13,YFYR14},
we introduced the multi-pomeron exchange potential (MPP) as a model 
of universal repulsions among three and four baryons on the basis 
of the Extended Soft Core (ESC) baryon-baryon interaction model 
developed by two of authors (T.R. and Y.Y.) and 
M.M. Nagels~\cite{ESC08,ESC08c1,ESC08c2,ESC08c3}.
Here, the Brueckner-Hartree-Fock (BHF) formalism was used
to derive the EoS of neutron-star matter including hyperons 
($\Lambda$ and $\Sigma^-$).

The most important point in such a work is how to determine the 
strength parameters related to for the stiffness of EoS:
In our approach, those of MPP were determined on the basis of 
terrestrial experiments, introducing no ad hoc parameter for the stiffness.
Usually, it is considered that some information on the 
incompressibility $K$ of high-density matter can be extracted from
analyses of central heavy-ion collisions in high energies. 
In the present, however, the results for the EoS still remain inconclusive.
On the other hand, in~\cite{YFYR13,YFYR14} we used the result that 
the TNR effect appeared in the experimental angular distributions of 
$^{16}$O+$^{16}$O elastic scattering ($E/A$=70 MeV), 
as shown in \cite{FSY,FSY14}:
Such a scattering phenomenon was analyzed successfully with
the complex G-matrix folding potentials derived from ESC
free-space $N\!N$ interactions including MPP contributions,
where the strengths of MPP are adjusted so as to reproduce 
the experimental data.

As well known, the nuclear saturation property of the density and 
energy per particle cannot be reproduced only with use of
two-body interactions. It is indispensable to take into account
the three-nucleon interaction composed of the attractive part 
(TNA) and the TNR~\cite{Panda81}.
When we introduce the MPP, the TNA is added phenomenologically
so as to reproduce the nuclear saturation property precisely.

%%%%%%%%revise%%%%%%%%%%%%%%%%%%%%%%%%%%%%%%%%%%%%%%%%%%%%%%%%%%%%%%%%
Our interaction model composed of ESC, MPP and TNA is extended to 
hyperon channels:
ESC gives potentials in $S=-1$ ($\Lambda\!N$, $\Sigma\!N$) and 
$S=-2$ ($\Xi\!N$, $\Lambda\!\Lambda$ and $\Lambda\!\Sigma$) channels. 
MPP is universal in all $B\!B$ channels according to its modeling.
TNA is given phenomenologically in $N\!N$ channels, which should 
be taken to reproduce hypernuclear data in hyperonic channels.
Thus, we have a three-baryon attraction (TBA).
%Thus, our baryon-baryon ($B\!B$) interaction is composed of 
%ESC, MPP and TBA. 
%
The decisive role for stiffness of EoS and neutron-star mass
is played by the MPP part. The existence of high-density strong 
repulsions in hyperon channels is an assumption to avoid the
softening of EoS by hyperon mixing in neutron-star matter.
Such an assumption is modeled by MPP in a clear-cut way.
Then, the strength of MPP should be determined so as to be 
consistent with the terrestrial experimental data.

%The phenomenological parts TNA in hyperonic channels
%Thus, MPP is universal in these channels.
%The phenomenological parts TNA in hyperonic channels
%are taken to reproduce hypernuclear data, being
%considered as a three-baryon attraction (TBA).
%%%%%%%%%%%%%%%%%%%%%%%%%%%%%%%%%%%%%%%%%%%%%%%%%%%%%%

Using our $B\!B$ interaction model (ESC+MPP+TBA),
we derive the EoS of $\beta$-stable neutron-star matter composed of
neutrons ($n$), protons ($p$), electrons ($e^-$), muons ($\mu^-$)
and hyperons ($\Lambda$, $\Sigma^-$, $\Xi^-$), and solve the 
Tolmann-Oppenheimer-Volkoff (TOV) equation for the hydrostatic 
structure to obtain mass-radius relations of neutron stars.

The first step in this paper is to find the parameter sets for
MPP and TNA in $N\!N$ channels on the basis of the G-matrix
calculations for nucleon matter. The nucleus-nucleus potentials
are derived by folding the G-matrix interactions consistent with
nuclear saturation properties. The MPP parameters are chosen
so as to reproduce the $^{16}$O+$^{16}$O scattering data 
at $E/A$=70 MeV. 
%Though such a task was already given in refs~\cite{YFYR13,YFYR14}, 
%another parameter set is added to give a stiffer EoS. 
In the nuclear matter calculations, 
a three-body interaction is replaced approximately to an 
effective two-body interaction by integrating out a third particle. 
%Then, in general it is 
%necessary to take into account short range correlations between
%two particles. This correlation effect, not considered in
%refs~\cite{YFYR13,YFYR14}, is estimated in this work.
%Furthermore, we investigate the other types of TNR model
%the Urbana UIX~\cite{UIX}. 
%and the string-junction model~\cite{SJM}.

The second step is to derive the EoS for neutron-star matter
including hyperons ($\Lambda$, $\Sigma^-$), and to obtain the 
mass-radius relations of neutron stars by solving the TOV equation.
Though the calculations were performed in \cite{YFYR14}, 
the obtained stiffness of the EoS was found to be slightly 
over-estimated due to some insufficiency in numerical calculations.
The correct results are given in this work.

The third step is to study the effect of $\Xi^-$ mixing,
which was not taken into account in \cite{YFYR14}.
In this relation, it is important that recently the event of 
$\Xi^-$ bound state was found in emulsion and the
$\Xi-$ binding energy $B_{\Xi^-}$ was extracted~\cite{Nakazawa}.
The $S=-2$ sector of ESC reproduces nicely the observed value
of $B_{\Xi^-}$. Then, we derive the EoS of neutron-star matter
including $\Xi^-$ together with $\Lambda$ and $\Sigma^-$,
and evaluate the effect of $\Xi^-$ mixing to the mass-radius
relations of neutron stars.

The paper is organized as follows:
In section \ref{sec:1}, the strengths of MPP and TBA in
nucleonic channels are determined to reproduce the angular
distributions of $^{16}$O+$^{16}$O scattering at $E/A$=70 MeV
and the nuclear saturation property.
The EoS is derived from a mixed matter of $n$, $p$, $e$ and $\mu$ 
in chemical equilibrium, and the mass-radius relations of
neutron stars are obtained.
In section \ref{sec:2}, the EoS is derived from a baryonic
matter including not only nucleons but also hyperons ($\Lambda$
and $\Sigma^-$). In spite of substantial softening of the EoS
by hyperon mixing, the resultant values of maximum masses of
neutron stars reach to $2M_{\odot}$ owing to the contributions
of quartic pomeron exchange terms.
The effects of $\Xi^-$ mixing are also investigated.
The conclusions of this work are given in section \ref{sec:3}.

\section{EoS and neutron-star mass}
\label{sec:1}

\subsection{Multi-pomeron repulsion}
\label{sec:11}

We start the $B\!B$ interaction model ESC,
where all available $N\!N$-, $Y\!N$-, and $Y\!Y$-data are fitted 
simultaneously with single sets of meson parameters.
Here, two-meson and meson-pair exchanges are taken into 
account explicitly and no effective boson is included differently 
from the usual one-boson exchange models. 
The latest version of ESC model is named 
as ESC08c~\cite{ESC08c1,ESC08c2,ESC08c3}.
Hereafter, ESC means this version.

As a model of universal TBR, we introduce
the multi-pomeron exchange potential (MPP) \cite{YFYR13,YFYR14}
consistently with the ESC modeling:
Generally, the N-body local potential by pomeron exchange is
%\begin{subequations}
\begin{eqnarray}
%&& V({\bf x}'_1, ..., {\bf x}'_N; {\bf x}_1, ... , {\bf x}_N) \equiv
% W^{(N)}({\bf x}_1, ..., {\bf x}_N)\ \Pi_{i=1}^N \delta({\bf x}'_i-{\bf x}_i), \\
 && W^{(N)}({\bf x}_1, ..., {\bf x}_N) = g_P^{(N)} g_P^N\ \left\{
\int\frac{d^3k_i}{(2\pi)^3} e^{-i{\bf k}_i\cdot{\bf x}_i}\right\}
\cdot\nonumber\\ && \times (2\pi)^3\delta(\sum_{i=1}^N {\bf k}_i)
 \Pi_{i=1}^N \left[\exp\left(-{\bf k}_i^2\right)\right]\cdot {\cal M}^{4-3N} ,
\label{eq:1}
\end{eqnarray}
%\end{subequations}
where the (low-energy) pomeron propagator is the same as used in the
two-body pomeron potential.
Since the pomeron is an SU(3)-singlet, MPP's work universally
among baryons. 
The effective two-body potential in a baryonic medium is obtained
by integrating over the coordinates ${\bf x}_3,..., {\bf x}_N$. 
This gives 
\begin{eqnarray}
&& V_{eff}^{(N)}({\bf x}_1,{\bf x}_2) 
 = \rho_{}^{N-2} 
 \int\!\! d^3\!x_3 ... \int\!\! d^3\!x_N\ 
 W^{(N)}({\bf x}_1,{\bf x}_2, ..., {\bf x}_N)
 \nonumber \\
&& 
=g_P^{(N)} g_P^N\frac{\rho^{N-2}}{{\cal M}^{3N-4}}\cdot
 \frac{1}{\pi\sqrt{\pi}} \left(\frac{m_P}{\sqrt{2}}\right)^3\
 \exp\left(-\frac{1}{2}m_P^2 r_{12}^2\right).
\label{eq:2}
\end{eqnarray}
%
%Since the pomeron is an SU(3)-singlet, MPP's universally
%among baryons. 
%in a nuclear medium 
%leads to a density-dependent universal repulsion~\cite{NYT}.
%%%%%%%%%%%%%%%%%%%%%%%%%%%%%%%%%%%
We assume that
the dominant mechanism is triple and quartic pomeron exchange.
%We restrict ourselves here to the triple and quartic pomeron couplings.
The values of the two-body pomeron strength $g_P$ and 
the pomeron mass $m_P$ are the same as those in ESC.
A scale mass ${\cal M}$ is taken as a proton mass.

%The density($\rho$)-dependent two-body potentials in a baryonic medium are 
%obtained by integrating over coordinates of third (and fourth)
%particles in the three-body (and four-body) potentials as follows:
%\begin{eqnarray}
%V^{(3)}_{eff}(r) 
%&=& g_P^{(3)} (g_P)^3 \frac{\rho}{{\cal M}^5} F(r) \ ,
%\\
%V^{(4)}_{eff}(r) 
%&=& g_P^{(4)} (g_P)^4 \frac{\rho^2}{{\cal M}^8} F(r) \ ,
%\\
%F(r) &=& \frac{1}{4\pi} \frac{4}{\sqrt{\pi}}
%\left(\frac{m_P}{\sqrt{2}}\right)^3
%\exp\left(-\frac12 m_P^2 r^2 \right) \ .
%\label{eq:3}
%\end{eqnarray}

In order to reproduce the nuclear saturation property,
an adequate form of TNA must be added to ESC together 
with MPP. Here, we introduce TNA phenomenologically 
as a density-dependent two-body interaction
\begin{eqnarray}
%V_{A}(r;\rho)= V_{0}\, \exp(-(r/2.0)^2)\, \rho\, 
%\exp(-\eta\rho)\ .
V_{A}(r;\rho)= V_{0}\, \exp(-(r/2.0)^2)\, \rho\, 
\exp(-\eta \rho)\, (1+P_r)/2 \ ,
\label{eq:4}
\end{eqnarray}
$P_r$ being a space-exchange operator. 
Here, the functional form is taken to be similar to 
the TNA part given in \cite{Panda81}. 
$V_{0}$ and $\eta$ are treated as adjustable parameters.
$V_{A}(r;\rho)$ works only in even states due to the $(1+P_r)$ factor.
This assumption is needed to reproduce the $^{16}$O$+^{16}$O potential
at $E/A=70$ MeV and nuclear-matter energy consistently \cite{YFYR14}.

\subsection{G-matrix calculation and determination of MPP strength}
\label{sec:12}

The lowest-order Bruckner G-matrix calculations 
with the continuous (CON) choice for intermediate single particle potentials
were shown to simulate well the results including higher hole-line contributions
up to $3\sim 4$ $\rho_0$~\cite{Baldo98,Baldo02}, $\rho_0$ being normal density.
%Here, the Bruckner G-matrix theory is considered a good starting point 
%for studies of many-body systems on the basis of
%free-space baryon-baryon interaction models. 
Here, G-matrix calculations are performed in nuclear matter, 
and G-matrix interactions are represented in coordinate space
to construct nucleus-nucleus folding potentials~\cite{FSY}. 

\begin{table}
\caption{Values of parameters included in MPP and TNA.}
\label{tab:1}       % Give a unique label
\begin{tabular}{lcccc}
\hline\noalign{\smallskip}
& $g_P^{(3)}$ & $g_P^{(4)}$ & $V_0$ & $\eta$ \\ 
\noalign{\smallskip}\hline\noalign{\smallskip}
MPa     & 2.34 & 30.0  & $-$32.8 &  3.5  \\
MPa$^+$ & 1.31 & 80.0  & $-$21.6 &  1.0  \\
MPb     & 2.94 & 0.0   & $-$45.0 &  5.4  \\
%cMPb    & 4.53 & 0.0   & $-$45.0 &  5.4  \\
\noalign{\smallskip}\hline
\end{tabular}
%\vspace*{5cm}  % with the correct table height
\end{table}

In the same way as \cite{YFYR13,YFYR14},
the analyses for the $^{16}$O$+^{16}$O elastic scattering at 
an incident energy per nucleon $E_{in}/A=70$ MeV are performed.
The MPP strengths ($g_P^{(3)}$ and $g_P^{(4)}$) and the TNA parameters 
($V_{0}$ and $\eta$) are determined to reproduce the scattering data 
using the G-matrix folding potential derived from ESC+MPP+TNA
and nuclear-matter energy at saturation density.
The calculated angular distributions of $^{16}$O$+^{16}$O scattering
are insensitive to the TNA parameters. Namely, it is essential in 
our approach that they are substantially determined by the MPP 
repulsive contributions in high density region~\cite{FSY14}.
On the other hand, they are not so dependent on a ratio of 
contributions of triple and quartic pomeron exchanges, and we can find
various combinations of $g_P^{(3)}$ and $g_P^{(4)}$ reproducing 
the data equally well.

The chosen parameter sets are listed in Table \ref{tab:1},
%where G-matrix calculations are performed without taking account
%of correlation functions in Eq.(\ref{eq:3}).
where MPa and MPb sets were used also in \cite{YFYR14}.
As stated in \cite{YFYR13}, the analyses of the experimental 
cross sections of the process $pp \rightarrow pX$ gives rise to
rough estimations $g_P^{(3)}= 1.95 \sim 2.6 $ and
$g_P^{(4)} = 33 \sim 228$ \cite{Kai74,Bron77}. 
In the case of MPa, the $g_P^{(3)}$ value is within the estimated 
range, and the $g_P^{(4)}$ is of a permissible minimum value.
MPa$^+$ is specified by having a fairly larger value of $g_P^{(4)}$ 
than MPa. On the other hand, MPb has no quartic component
($g_P^{(4)}=0$).
As found in Eq.(\ref{eq:2}), the contributions from triple and
quartic components are proportional to $\rho$ and $\rho^2$,
respectively. Therefore, the latter contribution play a remarkable
role to stiffen the EoS in high density region. 

\begin{figure}
%\resizebox{0.75\textwidth}{!}{%
\resizebox{0.5\textwidth}{!}{%
  \includegraphics{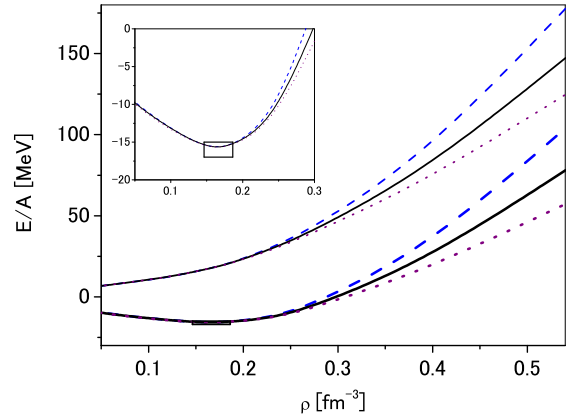}
}
\caption{Energy per particle ($E/A$) as a function of 
nucleon density $\rho$. Upper (lower) curves are for neutron 
matter (symmetric matter). Solid, dashed and dotted curves are 
for MPa, MPa$^+$ and MPb, respectively.
%Solid curves are obtained with MPa$^+$, MPa and MPb.
The box shows the empirical value. The inset shows a zoom
of the region around the saturation point.
}
\label{saturation}       % Give a unique label
\end{figure}

%%%%%%%%%%%%%%%%%%%%%%%%%%%%%%%%%%%%%%%%%%%%
In Fig.~\ref{saturation}, 
we show the energy curves of symmetric nuclear matter (lower curves) 
and neutron matter (upper curves), namely binding energy per nucleon 
($E/A$) as a function of density $\rho$. 
%They are obtained from G-matrix 
%calculations with MPa, MPa$^+$, MPb and cMPb. 
%Here, G-matrices are calculated with the CON choice for intermediate 
%nucleon spectra. 
%The dotted curves are obtained only with the two-body interaction ESC08c. 
%The saturation point in symmetric nuclear matter 
%is found to deviate substantially from the box. On the other hand,
%Top, middle and bottom solid curves are obtained by 
%MPa$^+$, MPa and MPb sets, respectively.
%Dashed curves are by cMPb with the correlation effect, 
%which is found to be similar to the curves for MPb.
%Dotted curved are by cUIX, being very similar to those for cMPb.
Solid, dashed and dotted curves are obtained by MPa, MPa$^+$ 
and MPb sets, respectively.
The box in the figure shows the area where nuclear 
saturation is expected to occur empirically.
Then, saturation densities and minimum values of $E/A$ curves
by these sets turn out to be nicely close to the empirical value.
In order to derive the compressibility $K$, the saturation curve 
is fitted by a function $E/A=a\rho + b\rho^\gamma$
at 0.07$<\rho <$ 0.4 fm$^{-3}$.
The obtained values of $K$ are 283, 313 and 254 MeV
for MPa, MPa$^+$ and MPb, respectively, where the saturation points
are the same value of $\rho_0=0.154$ fm$^{-3}$.
 
%%%%%%%%%%%%%%%%%%%%%%%%%%%%%%%%%%%%%%%%%%%%

\begin{figure}
%\resizebox{0.75\textwidth}{!}{%
\resizebox{0.4\textwidth}{!}{%
  \includegraphics{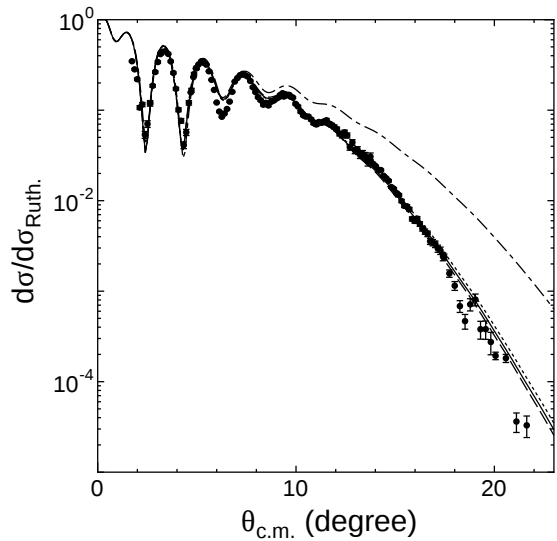}
}
%\vspace{3cm}
\vspace{0.5cm}
\caption{Differential cross sections for $^{16}$O+$^{16}$O elastic 
scattering at $E/A=70$ MeV calculated with the G-matrix folding potentials.
Solid, dashed and dotted curves are for MPa, MPa$^+$ and MPb, 
respectively. Dot-dashed curve is for ESC.}
\label{xsO16O16}       % Give a unique label
\end{figure}

\begin{figure}
%\resizebox{0.75\textwidth}{!}{%
\resizebox{0.4\textwidth}{!}{%
  \includegraphics{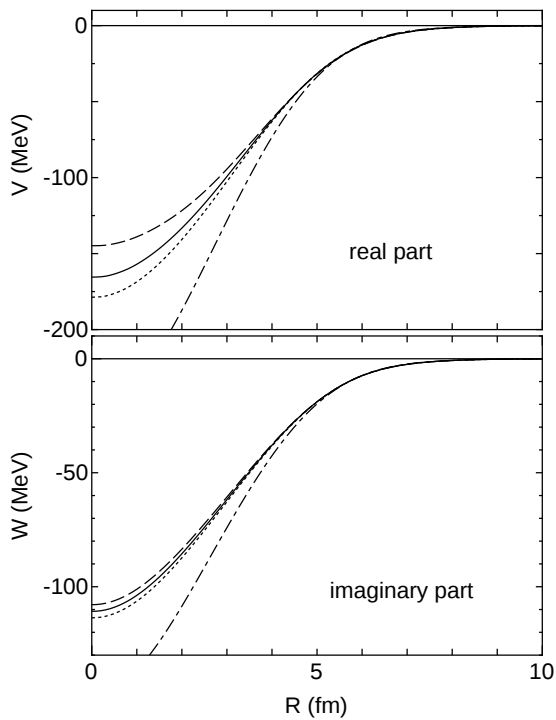}
}
%\vspace{2cm}
\vspace{0.5cm}
\caption{Double-folding potentials for $^{16}$O+$^{16}$O elastic 
scattering at $E/A=70$ MeV. Solid, dashed and dotted curves are for
MPa, MPa$^+$ and MPb, respectively. Dot-dashed curve is for ESC.
}
\label{potO16O16}       % Give a unique label
\end{figure}

In Fig.\ref{xsO16O16}, the calculated results of the differential cross sections 
for the $^{16}$O+$^{16}$O elastic scattering at $E/A=70$ MeV are compared with
the experimental data~\cite{Nuoffer}. 
The corresponding $^{16}$O+$^{16}$O double-folding potentials are shown
in Fig.\ref{potO16O16}.
Here, the dot-dashed curves are obtained from ESC, and the angular distribution
deviates substantially from the data.
Solid, dashed and dotted curves are for MPa, MPa$^+$ and MPb, respectively,
which reproduce nicely the experimental data.
%The result for cMPb is almost the same as that for MPb.
Though a reduction factors $N_W$ is often multiplied on the imaginary part
in the folding model analyses~\cite{FSY},
such a reduction factor is not needed in the present cases.
%Though the curve for cUIX is slightly below those for MPP models,
%it becomes similar by multiplying $N_W=0.95$ on the imaginary part.
%though the fitting by MPc seems to be slightly worse than MPa/b.
In the double-folding model analyses, the most important is the 
validity of the frozen-density approximation (FDA).
%Owing to the FDA, the MPP repulsion in the density region over the
%normal density contributes to folding potentials.
%Such an effect can be seen in Fig.\ref{potO16O16}, where the potentials
%for MPa/b/c are remarkably shallower than that for ESC.
As shown in \cite{FSY14}, MPP contributions from the density region 
higher than the normal density are decisively important for resultant 
angular distributions. Namely, valuable information of the EoS in 
high-density region can be obtained from double-folding potentials 
with FDA.
The effect to include the quartic pomeron coupling has to appear in 
the difference between results for MPa/MPa$^+$ and MPb, but 
no meaningful effect can be found in the present analyses for 
nucleus-nucleus scattering.

It is instructive to compare our MPP with another model, for instance, 
the short-range repulsive term in the Urbana model IX (UIX)~\cite{UIX}.
%$V_{ijk}^R(UIX)$~\cite{UIX}.
%Then, the MPP part in cMPb is replaced by $V_{ijk}^R(UIX)$, and 
%$V_0$ in TNA is taken as $-66$ MeV to reproduce the saturation energy.
%This set is denoted by cUIX.
Because the strength of UIX is determined on the basis of 
variational calculations for nuclear systems,
two-body correlations should be taken into account in deriving 
an effective two-body potential from a three-body potential
for use of UIX in our G-matrix calculations.
%
%$As seen above, an effective two-body potential is derived by 
%integrating over coordinates other than respective two particles.
%Generally, in this procedure it is necessary to take account of
%two-body correlations: When a two-body correlation between $(i,j)$ 
%particles is given by a correlation function $f({\bf x}_i,{\bf x}_j)$,
%an effective two-body potential obtained from a three-body potential
%is represented as
%\begin{eqnarray}
%  \tilde V_{eff}^{(2)}({\bf x}_1,{\bf x}_2) 
% = &\rho&
% \int\!\! d^3\!x_3\ 
% W^{(3)}({\bf x}_1,{\bf x}_2,{\bf x}_3)
% \nonumber \\
% &\times&\ f^2({\bf x}_1,{\bf x}_3) \ f^2({\bf x}_3,{\bf x}_2) \ .
%\label{eq:3}
%\end{eqnarray}
We estimate the effect of two-body correlations in the case of MPb 
with no quartic component, where correlation functions are extracted 
from the solutions of G-matrix equation in nuclear matter.
%For simplicity, we use correlation functions obtained from 
%a statistically-weighted average of functions in 
%$^1S_0$ and $^3P_J$ states \cite{SJM}.
The G-matrix results from MPb without correlations
are found to be very similar to the results including correlations,
if the $g_P^{(3)}$ value in MPb is multiplied by about 1.5.
Then, in comparison with the calculations including UIX
instead of the MPP part in MPb, it turns out that 
the G-matrices in both cases give rise to similar results.
Namely, the strengths of three-body repulsions of MPb and UIX 
are rather similar to each other.
%This set is given as cMPb in Table \ref{tab:1}.

\subsection{neutron star mass}
\label{sec:13}

From the G-matrix calculations, we obtain the energy per nucleon $E/A$ 
as a function of density $\rho$.
The $E/A$ curve in nuclear matter with neutron density $\rho_n$ 
and proton density $\rho_p$ is parameterized as
\begin{eqnarray}
E/A = (1-\beta) (a_0 \rho + b_0 \rho^{c_0})
            +\beta (a_1 \rho+ b_1 \rho^{c_1}) 
\label{EA}
\end{eqnarray}
with $\beta=1-2x$, $x=\rho_p/\rho$ and $\rho=\rho_n+\rho_p$.

\begin{table}
\caption{Parameters in Eq.(\ref{EA})}
\label{tab:2}      
\begin{tabular}{lcccccc}
\hline\noalign{\smallskip}
& $a_0$ & $b_0$ & $c_0$ & $a_1$ & $b_1$ & $c_1$ \\ 
\noalign{\smallskip}\hline\noalign{\smallskip}
MPa     & $-234.8$ & 643.8 & 1.86 & 66.41 & 490.1 & 2.40 \\
MPa$^+$ & $-186.1$ & 814.9 & 2.23 & 76.36 & 808.7 & 2.83 \\
MPb     & $-376.4$ & 639.0 & 1.47 & 36.46 & 334.8 & 1.89 \\
\noalign{\smallskip}\hline
\end{tabular}
%\vspace*{5cm}  % with the correct table height
\end{table}

\begin{figure}
%\resizebox{0.75\textwidth}{!}{%
\resizebox{0.5\textwidth}{!}{%
\includegraphics{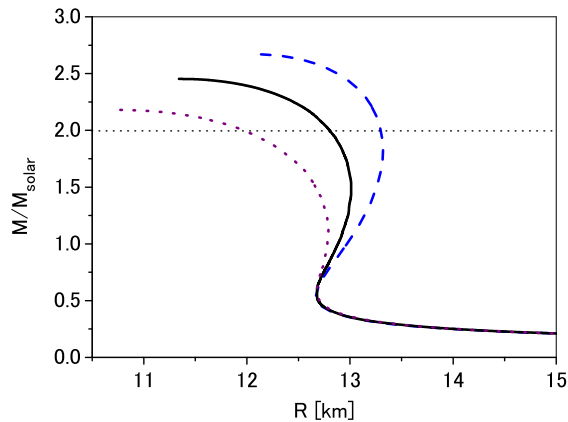}
}
\caption{
Neutron-star masses as a function of the radius $R$.
Solid, dashed and dotted curves are for MPa, MPa$^+$ and MPb,
respectively.
%Dashed and dotted curves are for cMPb and cUIX, respectively.
}
\label{MRnuc}
\end{figure}

Energy density, chemical potential and pressure are obtained 
from Eq.(\ref{EA}) as a function of $\rho$ and $\beta$. 
Assuming a mixed matter of $n$, $p$, $e^-$ and $\mu^-$
%neutrons, protons, electrons and muons 
in chemical equilibrium, we solve the TOV equation for 
the hydrostatic structure of a spherical nonrotating star. 
The obtained mass-radius relations of neutron stars are 
demonstrated in Fig.\ref{MRnuc}. Solid, dashed and dotted 
curves are for MPa, MPa$^+$ and MPb, respectively.
%As shown by solid curves for MPa, MPa$^+$ and MPb,
The EoS's in these cases are found to be stiff enough 
to give 2$M_{\odot}$. 
The difference between MPa (MPa$^+$) and MPb is due to the 
quartic-pomeron exchange term included in the formers. 
The strengths of the effective two-body interaction derived 
from  quartic-pomeron exchanges are proportional 
to $\rho^2$, and the contribution become sizeable in the 
high-density region, making the maximum mass large.
The above differences appear significantly in the inner regions 
%considered to the significant difference among 
of $^{16}$O+$^{16}$O double-folding potentials 
in the inner region, though they cannot be seen 
in the cross sections Fig.\ref{xsO16O16}. 
%The quartic-pomeron exchange contribution cannot be taken 
%out by present analyses of nucleus-nucleus scattering data.

%The result for cMPb is shown by the dashed curve, where
%the obtained masses are only slightly smaller than those
%for MPb. The dotted curve for cUIX is very similar to
%that for cMPb. This means that the three-body repulsions
%included in both models are of similar strengths. 

\section{hyperon mixing in neutron stars}
\label{sec:2}

\subsection{$Y\!N$ interaction}
\label{sec:21}

ESC gives potentials in $S=-1$ ($\Lambda\!N$, $\Sigma\!N$) and 
$S=-2$ ($\Xi\!N$, $\Lambda\!\Lambda$ and $\Lambda\!\Sigma$) channels,
being designed consistently with various data of $Y\!N$ scattering
and hypernuclei. Then,
the most important is to test the MPP+TBA parts in channels 
including hyperons. While MPP is defined universally in all
baryon channel, TBA is introduced phenomenologically in nucleon
channels, and not defined in $Y\!N$ channels.
%%%%%%%%revise%%%%%%%%%%%%%%%%%%%%%%%%%%%%%%%%%%%%%%%%%%%%%%%%%%%%%%%%%%
Then, we should determine the strength of TBA in each $Y\!N$ channel
so as to reproduce the related hypernuclear data.
Fortunately, the present version of ESC reproduces well the basic 
features of $S=-1$ and $S-2$ systems obtained from experimental data.
%Then, it is quite important that ESC reproduces well the
%basic features of $S=-1$ and $S-2$ systems.
This means that MPP contributions in $YN$ channels
as well as those in $NN$ channels are canceled substantially 
by TBA contributions at normal density region. 
Then, it is confirmed that the hypernuclear data are reproduced
by ESC+MPP+TBA as well as ESC by choosing TBA in a $Y\!N$ channel 
equally to that in a $N\!N$ channel.
%Here, we assume simply that TBA in a $Y\!N$ channel is equal to 
%that in a $N\!N$ channel. 
One should notice, however, that the validity of this simple choice 
of TBA is rather accidental. There still remains ambiguities
in the parameter fitting of ESC due to lack of experimental data
in $Y\!N$ channels. When another parameter set of ESC is used,
it is likely that TBA in a $Y\!N$ channel is different from TNA
in a $N\!N$ channel.

In the case of $\Lambda\!N$ case,
such an assumption can be tested in detail by using the 
experimental data of $\Lambda$ hypernuclei:
We calculate $\Lambda\!N$ G-matrices in symmetric nuclear matter 
including a single $\Lambda$ hyperon.
In Table~\ref{tab:3} we show the potential energies $U_\Lambda$
for a zero-momentum $\Lambda$ and their partial-wave contributions
in $^1S_0$, $^3S_1$, $P$ and $D$ states 
at normal density $\rho_0$ ($k_F$=1.35 fm$^{-1}$).
It is noted that reasonable $\Lambda$ binding energies are obtained
in the cases of MPa, MPa$^+$, MPb and ESC.
The former three results including MPP+TBA are found to be
similar to the ESC result, because MPP and TBA contributions
are cancelled out substantially in normal density region
in spite of remarkable difference in higher density region.
%%%%%%%revise%%%%%%%%%%%%%%%%%%%%%%%%%%%%%%%%%%%%%%%%%%%%%%
%It should be noted that these values of $U_\Lambda(\rho_0)$ 
%are not compared directly to the depth $U_{WS}\sim -30$ MeV of 
%the $\Lambda$ Woods-Saxon (WS) potential suitable to the data of 
%$\Lambda$ hypernuclei.
One should be careful for comparing these values of 
$U_\Lambda(\rho_0)$ with the depth $U_{WS}\sim -30$ MeV of 
the $\Lambda$ Woods-Saxon (WS) potential suitable to the data of 
$\Lambda$ hypernuclei.
In the cases of using the Skyrme-type $\Lambda\!N$ interactions
\cite{Skyrme} or the $\Lambda$ energy densities \cite{Schlze00}
in calculations of $B_\Lambda$ values in finite systems,
the derived $\Lambda$-nucleus potentials are more or less
similar to the WS form and the potential depths have
good correspondence to the $U_{WS}$ value. On the other hand,
in \cite{YFYR14} $\Lambda$ binding energies in finite systems were
calculated systematically with the $\Lambda$-nucleus folding potentials
derived from finite-range G-matrix interactions $G_{\Lambda\!N}(r;k_F)$ 
for MPa and ESC, and the experimental data were reproduced nicely. 
The similar result is obtained for MPb,
while that for MPa$^+$ is a little smaller than the data. 
In these cases, the forms of $\Lambda$-nucleus folding potentials
are considerably different from the WS form, and then
the potential depths are not simply compared with the WS one.
The reason why we use the G-matrix folding potentials
for hyperon-nucleus interactions is because nucleon-nucleus and 
nucleus-nucleus scattering phenomena are quite
successfully reproduced with G-matrix folding models.
In \cite{YFYR14}, such an analysis was used to extract
the three-body repulsive effect from the nucleus-nucleus
scattering observable. For consistency in our approach, we use 
the G-matrix folding procedures for hyperon-nucleus potentials.

%Anyway, the $\Lambda$-nucleus folding potential depends not only on the 
%strengths of $\Lambda N$ G-matrices but also on their density dependences.
%Thus, we can say that it is inadequate to consider the depth $U_{WS}$ 
%of the phenomenological Woods-Saxon potential of $\Lambda$ as the 
%$\Lambda$ potential depth in nuclear matter.

\begin{table}
\caption{Values of $U_\Lambda$ at normal density and partial wave
contributions for MPa, MPa$^+$, MPb and ESC (in MeV).
%from the G-matrix calculations. 
%with  CON prescriptions (in MeV).
Values specified by $P$ and $D$ give sum of $(S,J)$ contributions.
}
\label{tab:3}
\begin{tabular}{l|cccc|c}
\hline\noalign{\smallskip}
& $^1S_0$ & $^3S_1$ & $P$ & $D$ & $U_\Lambda$  \\
\noalign{\smallskip}\hline\noalign{\smallskip}
MPa    &$-$13.6& $-$25.9& 4.1 & $-$2.7 & $-$38.1 \\
MPa$^+$&$-$13.3& $-$25.1& 4.3 & $-$2.7 & $-$36.9 \\
MPb    &$-$13.6& $-$26.0& 4.1 & $-$2.7 & $-$38.3 \\
ESC    &$-$13.6& $-$25.3& 1.1 & $-$1.6 & $-$39.4 \\
\noalign{\smallskip}\hline
\end{tabular}
\end{table}

In Table~\ref{tab:4} we show the potential energies $U_\Sigma(\rho_0)$
for a zero-momentum $\Sigma$ and their partial-wave contributions 
%in $(T, ^{2S+1}L_J)$ states 
for MPa, MPa$^+$, MPb and ESC. 
As well as the cases of $U_\Lambda(\rho_0)$, the results for MPa, MPa$^+$ 
and MPb are similar to the ESC result because of cancellations of MPP 
and TBA contributions in normal density region.
It should be noted here that 
the strongly repulsive contributions in $T=3/2$ $^3S_1$  and
$T=1/2$ $^1S_0$ states are due to the Pauli-forbidden effects
in these states, being taken into account by strengthening the 
pomeron coupling in the ESC08 modeling.
Especially, $\Sigma^-$ potentials in neutron matter become 
strongly repulsive owing to $T=3/2$ $^3S_1$ contributions.
From the experimental data of $\Sigma^-$ hypernuclear production,
the $\Sigma$-nucleus potential is suggested to be strongly repulsive.
It was shown that the experimental $K^+$ spectra of 
$^{28}$Si$(\pi^-,K^+)$ reaction were reproduced using 
the repulsive Woods-Saxon (WS) potential with the strength 
$U_{WS}=20 \sim 30$ MeV~\cite{Harada05}.  
%%%%%%%%%%%%%%%%%%%%%revise%%%%%%%%%%%%%%%%%%%%%%%%%%%%%%
%In the same reason as the $\Lambda$ case, however, this value of 
%$U_{WS}$ cannot be corresponded directly to the $\Sigma$ potential
%depth $U_\Sigma(\rho_0)$ in nuclear matter.
%Though the values of $U_\Sigma(\rho_0)$ in Table~\ref{tab:4}
%are far smaller than that of $U_{WS}$, 
It is meaningful to compare the WS potential with the $\Sigma$ potential 
obtained from $\Sigma\!N$ G-matrices for the $\Sigma$ particle 
with positive energy in nuclear matter.
Then, the latter is found to be repulsive comparably to the former
in the $\Sigma^-$ energy region related to the above $(\pi^-,K^+)$ reaction. 
It is an interesting subject in future to analyze the $(\pi^-,K^+)$ reaction
data with use of the $\Sigma$-nucleus G-matrix folding potentials.
%the $(\pi^-,K^+)$ strength functions obtained from the G-matrix folding 
%potentials can be shown to be of more repulsive nature than that from
%the above WS potential.

\begin{table}
\caption{Values of $U_\Sigma$ at normal density and partial wave
contributions for MPa, MPa$^+$, MPb and ESC (in MeV).
Values specified by $P$ and $D$ give sum of $(S,J)$ contributions.
}
\label{tab:4}
\begin{tabular}{lr|rrrr|r}
\hline\noalign{\smallskip}
&& $^1S_0$ & $^3S_1$ &  $P$\ &  $D$ & $U_\Sigma$  \\
\noalign{\smallskip}\hline\noalign{\smallskip}
MPa     & $T=1/2$ &     10.7& $-$23.1& $-$1.4 & $-$1.0 &      \\
        & $T=3/2$ &  $-$13.3&    30.4&    0.1 & $-$0.9 &  1.5 \\
\noalign{\smallskip}
MPa$^+$ & $T=1/2$ &     10.6& $-$21.3& $-$1.4 & $-$1.0 &      \\
        & $T=3/2$ &  $-$13.3&    30.4&    0.2 & $-$0.9 &  3.4 \\
\noalign{\smallskip}
MPb     & $T=1/2$ &     10.6& $-$23.2& $-$1.4 & $-$1.0 &      \\
        & $T=3/2$ &  $-$13.3&    30.3&    0.1 & $-$0.9 &  1.3 \\
\noalign{\smallskip}
ESC     & $T=1/2$ &     10.9& $-$21.6& $-$2.5 & $-$0.7 &      \\
        & $T=3/2$ &  $-$13.5&    31.0& $-$2.1 & $-$0.2 &  1.3 \\
\noalign{\smallskip}\hline
\end{tabular}
\end{table}

\subsection{Hyperonic nuclear matter and EoS}
\label{sec:22}

We derive the EoS of baryonic matter composed of nucleons ($N=n,p$) 
and hyperons ($Y=\Lambda, \Sigma^-, \Xi^-$).
A single particle potential of $B$ particle in $B'$ matter is given by
\begin{eqnarray}
U_B(k)&=&\sum_{B'} U_{B}^{(B')}(k) 
\nonumber
\\
 &=& \sum_{B'} \sum_{k',k_F^{(B')}} \langle kk'|G_{BB',BB'}|kk'\rangle
\end{eqnarray}
with $B,B'=N,Y$. Here, spin isospin quantum numbers are implicit.
%
% revise2
The energy density is given by
\begin{eqnarray}
\varepsilon&=&
\varepsilon_{mass}+
\varepsilon_{kin}+\varepsilon_{pot} 
\nonumber
\\
&=& 2\sum_{B} \int_0^{k_F^B} \frac{d^3k}{(2\pi)^3}
\left\{ 
%M_B-M_n+
M_B+
\frac{\hbar^2 k^2}{2M_B}+\frac 12 U_B(k)\right\} 
\end{eqnarray}
Then, we have
$$\int_0^{k_F^B} \frac{k^2 dk}{\pi^2} U_B^{(B')}(k)=
\int_0^{k_F^{B'}} \frac{k^2 dk}{\pi^2} U_{B'}^{(B)}(k)$$. 

\noindent
Considering $\rho_B=\frac{(k_F^B)^3}{3\pi^2}$
\begin{eqnarray}
\frac{\partial}{\partial \rho_B}{\cal U}_B^{(B')}=
U_B^{(B')}(k_F^B)+ \int_0^{k_F^{B}} \frac{k^2 dk}{\pi^2} 
\frac{\partial U_{B}^{(B')}(k)}{\partial \rho_B}
\label{ypot}
\end{eqnarray}
The second term leads to the rearrangement contribution.

The baryon number density is given as $\rho=\sum_B \rho_B$,
$\rho_B$ being that for component $B$.
The chemical potentials $\mu_B$ and pressure $P$
are expressed as
\begin{eqnarray}
&&\mu_B = \frac{\partial \varepsilon}{\partial \rho_B} \ , 
\label{chem} \\
&& P = \rho^2 \frac{\partial (\varepsilon/\rho)}{\partial \rho_B}
 =\sum_B \mu_B \rho_B -\varepsilon \ .
\label{eq:press}
\end{eqnarray}

%%%%%%%%%%%%%%%%%%%%%%revise%%%%%%%%%%%%%%%%%%%%%%%%%%%
Here, the two approximations are made for the energy density:
(1) Hyperonic energy densities including $\Lambda$, $\Sigma^-$ and
$\Xi^-$ are obtained from calculations of $n+p+\Lambda$,
$n+p+\Sigma^-$ and $n+p+\Xi^-$ systems, respectively.
(2) The parabolic approximation is used to treat asymmetries 
between $n$ and $p$ in $n+p$ sectors.
% revise2
Then, the G-matrix equations are solved for $Y\!N$ pairs ($Y=\Lambda, \Sigma$)
specified by isospin quantum numbers, where the $\Lambda\!N$-$\Sigma\!N$ 
coupling terms are treated exactly.
The obtained isospin-represented $Y\!N$ G-matrices are transformed into 
those for $\Lambda n$, $\Lambda p$, $\Sigma^- n$ and $\Sigma^- p$.
The corresponding terms for $\Xi\!N$ pairs are done
approximately as mentioned later.

Calculated values of energy densities are fitted by
the following analytical parameterization:
\begin{eqnarray}
&& \varepsilon_{pot}(\rho_n,\rho_p,\rho_\Lambda,\rho_\Sigma,\rho_\Xi) 
= E_N \rho_N
\nonumber \\
&& \quad +(E_\Lambda +E_{\Lambda \Lambda}) \rho_\Lambda
+E_\Sigma \rho_\Sigma 
+E_\Xi \rho_\Xi 
\label{eq:a1}
\\
&& \quad E_z=(1-\beta)f_z^{(0)} + \beta f_z^{(1)}  \\
&& \qquad z= \Lambda, \Sigma, \Xi, \Lambda \Lambda
\nonumber
\end{eqnarray}
\noindent
where $\beta=(1-2x_p)^2$ with $x_p=\rho_p/\rho_N$ and
$\rho_N=\rho_n+\rho_p$.
$E_N$ means $E/A$ given by Eq.(\ref{EA}).
Expressions of $f_z^{(i)}$ with $i=0, 1$ are given as
\begin{eqnarray}
%&& f_N^{(i)}=a_N^{(i)} \rho_N +b_N^{(i)} \rho_N^{c_N^{(i)}}
%\\
&& f_y^{(i)}=A_y^{(i)} \rho_N +B_y^{(i)} \rho_N^{c_y^{(i)}}
\\
&& \quad A_y^{(i)}=a_{y0}^{(i)}+a_{y1}^{(i)}x_Y %+a_{y2}^{(i)}x_Y^2
\\
&& \quad B_y^{(i)}=b_{y0}^{(i)}+b_{y1}^{(i)}x_Y %+b_{y2}^{(i)}x_Y^2
\label{eq:a2}
\end{eqnarray}
where $x_Y=\rho_Y/\rho_N$ with $Y=$ $\Lambda$, $\Sigma$, $\Xi$,
and $y=$ $\Lambda$, $\Sigma$, $\Xi$, $\Lambda\!\Lambda$. Here,
$\Lambda$, $\Sigma$, $\Xi$, $\Lambda\!\Lambda$ denote
contributions from $N\!\Lambda$, $N\!\Sigma^-$, $N\!\Xi^-$ 
$\Lambda\!\Lambda$ interactions, respectively.
The values of fitted parameters are listed in Table \ref{param},
where $\Lambda\!\Lambda$ $(i=0)$ parts are omitted because of
their negligible effects in the EoS.
%%%%%%%%%%%%%%%%%%%%%%revise%%%%%%%%%%%%%%%%%%%%%%%%%%%%%
In this work, $\Sigma^-\!\Sigma^-$ and $\Xi^-\!\Xi^-$
interactions are not taken into account, for which 
there is no experimental information.
%%% revise2
%In our present results, the effects of $\Lambda\!\Lambda$ interaction
%are quite small. If $\Sigma^-\!\Sigma^-$ and $\Xi^-\!\Xi^-$ interactions
%are of the same order as the $\Lambda\!\Lambda$ interaction one,
%their effects also have to be small.

\begin{table}
\caption{Parameters of energy densities 
given by analytical forms Eqs.(13)$\sim$(15).}
\label{param}
\begin{tabular}{lrrrrr}
\hline\noalign{\smallskip}
& $a_{\Lambda 0}^{(0)}$ & $a_{\Lambda 1}^{(0)}$  
& $b_{\Lambda 0}^{(0)}$ & $b_{\Lambda 1}^{(0)}$ & $c_\Lambda^{(0)}$ \\
MPa     & $-$343.4 &    82.30 & 1224. & 1121. &  2.288 \\
MPa$^+$ & $-$273.0 & $-$73.75 & 1449. & 3582. &  2.695 \\
MPb     & $-$1191. & $-$449.2 & 1512. & 912.1 &  1.246 \\
\noalign{\smallskip}
& $a_{\Lambda 0}^{(1)}$ & $a_{\Lambda 1}^{(1)}$ 
& $b_{\Lambda 0}^{(1)}$ & $b_{\Lambda 1}^{(1)}$ & $c_\Lambda^{(1)}$ \\
MPa     & $-$208.4 & $-$66.90 & 1124. & 2204. &  2.555 \\
MPa$^+$ & $-$262.7 & $-$228.6 & 1184. & 2902. &  2.338 \\
MPb     & $-$1133. & $-$184.5 & 1540. & 576.8 &  1.273 \\
\noalign{\smallskip}
& $a_{\Sigma 0}^{(0)}$ & $a_{\Sigma 1}^{(0)}$  
& $b_{\Sigma 0}^{(0)}$ & $b_{\Sigma 1}^{(0)}$ & $c_\Sigma^{(0)}$ \\
MPa     & $-$252.8 & $-$261.7 & 699.2 & 992.8 &  1.591 \\
MPa$^+$ & $-$72.87 & $-$80.57 & 666.6 & 1531. &  2.254 \\
MPb     & $-$847.0 & $-$594.3 & 1174. & 930.3 &  1.184 \\
\noalign{\smallskip}
& $a_{\Sigma 0}^{(1)}$ & $a_{\Sigma 1}^{(1)}$  
& $b_{\Sigma 0}^{(1)}$ & $b_{\Sigma 1}^{(1)}$ & $c_\Sigma^{(1)}$ \\
MPa     &    105.3 &    104.1 & 945.5 & 1447. &  2.645 \\
MPa$^+$ &    102.8 &    39.74 & 967.0 & 2628. &  2.777 \\
MPb     & $-$84.92 & $-$30.89 & 687.3 & 557.6 &  1.613 \\
\noalign{\smallskip}
& $a_{\Lambda \Lambda 0}^{(1)}$ & $a_{\Lambda \Lambda 1}^{(1)}$  
& $b_{\Lambda \Lambda 0}^{(1)}$ & $b_{\Lambda \Lambda 1}^{(1)}$ 
& $c_{\Lambda \Lambda}^{(1)}$ \\
MPa     & $-$1.185 & $-$69.88 & $-$20.17 & 612.2 &  2.272 \\
MPa$^+$ & $-$2.294 & $-$33.30 & $-$44.02 & 1090. &  2.885 \\
MPb     &    7.449 & $-$256.9 & $-$232.6 & 2325. &  1.869 \\
\noalign{\smallskip}\hline
\end{tabular}
\end{table}

Let us derive the EoS of neutron-star matter composed of
$n$, $p$, $e^-$, $\mu^-$, $\Lambda$, $\Sigma^-$ and $\Xi^-$.
The equilibrium conditions are summarized as follows:
\noindent
(1) chemical equilibrium conditions,
\begin{eqnarray}
\label{eq:c1}
&& \mu_n = \mu_p+\mu_e \\
&& \mu_\mu = \mu_e \\
&& \mu_\Lambda = \mu_n \\
&& \mu_{\Sigma^-} =\mu_n + \mu_e \\
&& \mu_{\Xi^-} =\mu_n + \mu_e
\label{eq:c2}
\end{eqnarray}
\noindent
(2) charge neutrality,
\begin{eqnarray}
\rho_p = \rho_e +\rho_\mu+\rho_{\Sigma^-}+\rho_{\Xi^-}
\end{eqnarray}
\noindent
(3) baryon number conservation,
\begin{eqnarray}
\rho = \rho_n +\rho_p +\rho_\Lambda+\rho_{\Sigma^-}+\rho_{\Xi^-}
\label{eq:c3}
\end{eqnarray}

When the analytical expressions (\ref{eq:a1})$\sim$(\ref{eq:a2})
are substituted into the chemical potentials (\ref{chem}),
the chemical equilibrium conditions (\ref{eq:c1})$\sim$(\ref{eq:c2})
are represented as equations for densities $\rho_a$
($a=$ $n$, $p$, $e^-$, $\mu^-$, $\Lambda$, $\Sigma^-$, $\Xi^-$).
Then, equations (\ref{eq:c1})$\sim$(\ref{eq:c3})
can be solved iteratively.
%$\Xi^-$ mixing is studied in the next subsection.

\begin{figure}
%\resizebox{0.75\textwidth}{!}{%
\resizebox{0.5\textwidth}{!}{%
\includegraphics{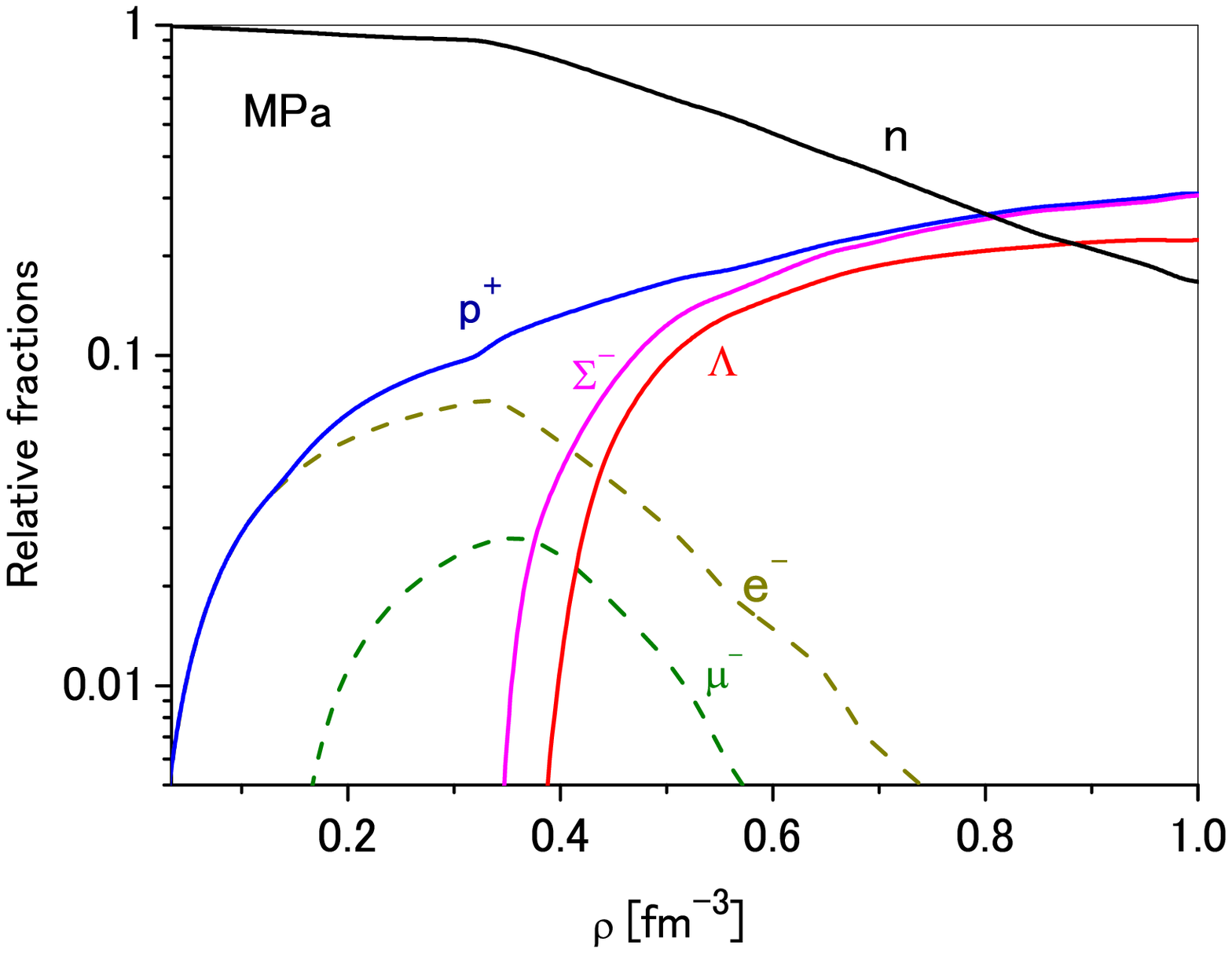}
}
\caption{Composition of hyperonic neutron-star matter 
for MPa.}
\label{chem1}
\end{figure}

\begin{figure}
%\resizebox{0.75\textwidth}{!}{%
\resizebox{0.5\textwidth}{!}{%
\includegraphics{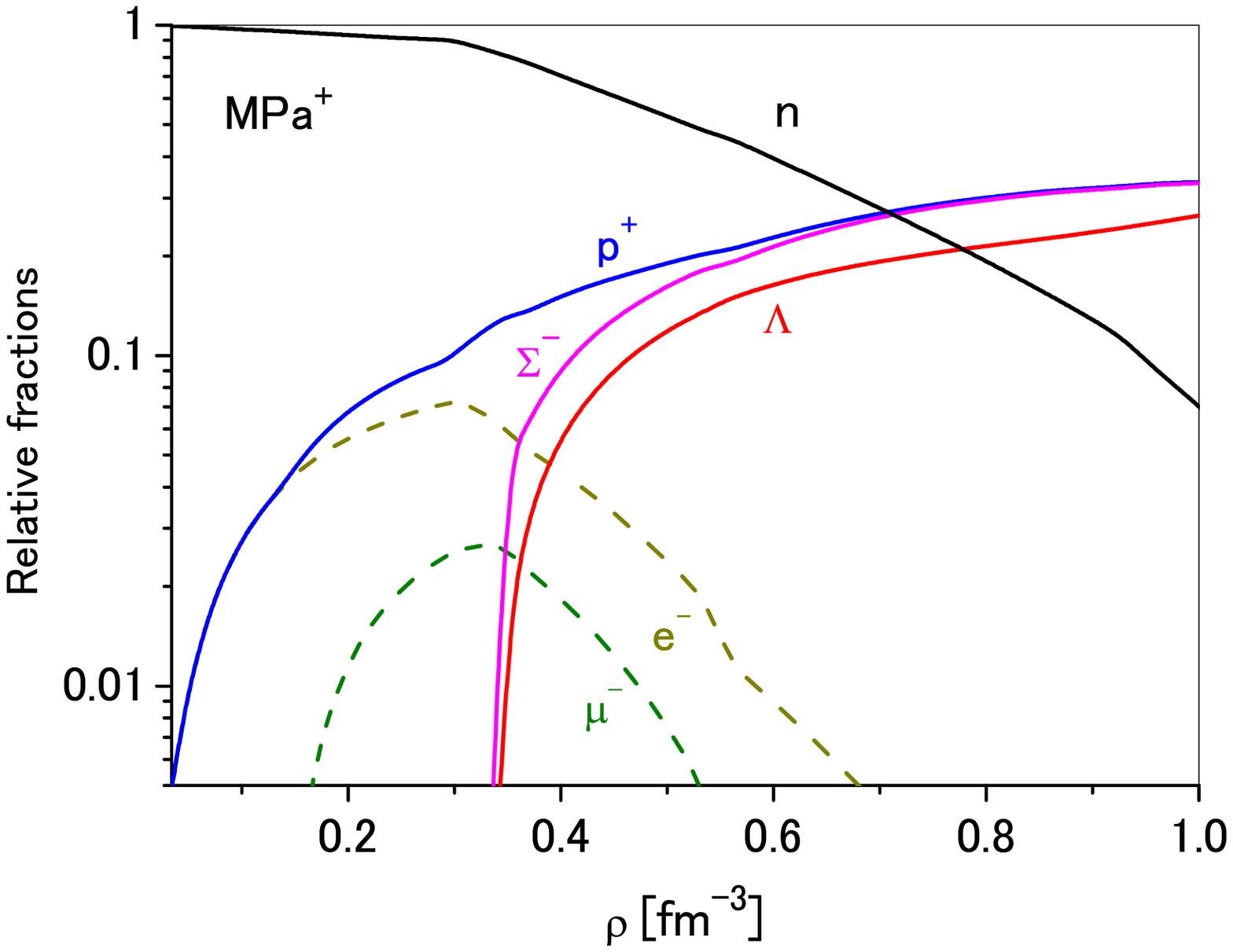}
}
\caption{Composition of hyperonic neutron-star matter 
for MPa$^+$.}
\label{chem2}
\end{figure}

\begin{figure}
%\resizebox{0.75\textwidth}{!}{%
\resizebox{0.5\textwidth}{!}{%
\includegraphics{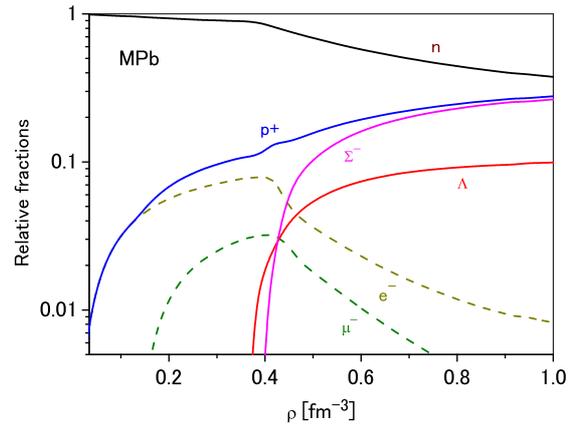}
}
\caption{Composition of hyperonic neutron-star matter 
for MPb.}
\label{chem3}
\end{figure}

In this subsection, let us show the results obtained without 
$\Xi^-$ mixing.
In Fig.~\ref{chem1}, Fig.~\ref{chem2} and Fig.~\ref{chem3},
the matter compositions are shown in the cases of MPa, MPa$^+$
and MPb, respectively. 
As the MPP repulsions become strong from MPb to MPa$^+$,
hyperon components become large.
Increasing of hyperon components are found to be  covered 
by decreasing of components of $n$, $e^-$ and $\mu^-$.

\begin{figure}
%\resizebox{0.75\textwidth}{!}{%
\resizebox{0.5\textwidth}{!}{%
\includegraphics{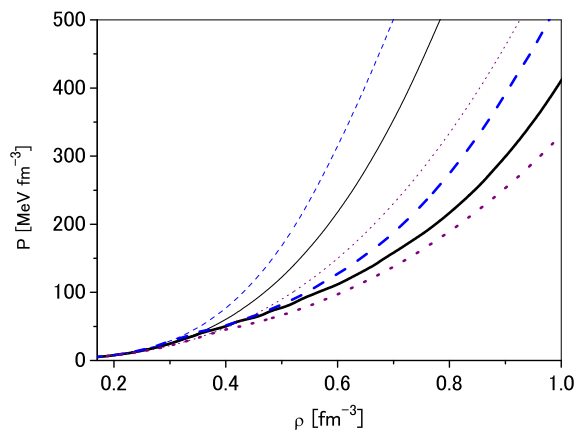}
}
\caption{Pressure $P$ as a function of baryon density $\rho$.
Thick (thin) curves are with (without) hyperon mixing.
Solid, dashed and dotted curves are for MPa, MPa$^+$ and MPb.
}
\label{press}
\end{figure}

Pressures (\ref{eq:press}) are obtained from 
determined values of densities and chemical potentials. 
In Fig.~\ref{press}, the calculated values of 
pressure $P$ are drawn as a function of 
%energy density $\varepsilon$ and 
baryon density $\rho$.
Thick (thin) curves are with (without) hyperon mixing.
Solid, dashed and dotted curves are for MPa, MPa$^+$ and MPb.
%Solid and dotted curves are with and 
%without hyperon mixing, respectively. 
%Top, middle and bottom solid (dashed) curves are
%with (without) hyperon mixing 
%in the case of MPa$^+$, MPa and MPb, respectively. 

\begin{figure}
%\resizebox{0.75\textwidth}{!}{%
\resizebox{0.5\textwidth}{!}{%
\includegraphics{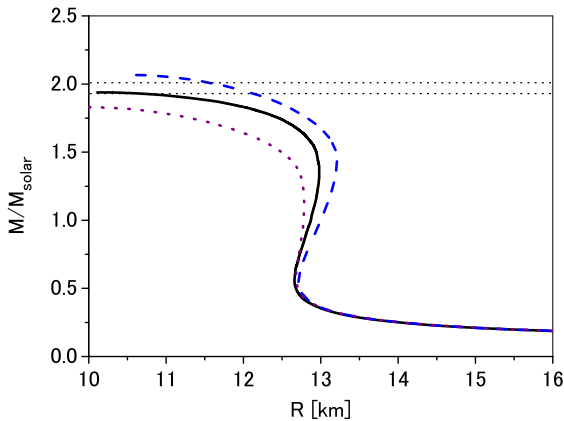}
}
\caption{Neutron-star masses as a function of the radius $R$.
Solid, dashed and dotted curves are for MPa, MPa$^+$ and MPb. 
Two dotted lines show the observed 
mass $(1.97\pm0.04)M_{\odot}$ of J1614-2230.
}
\label{MRhyp}
\end{figure}

\begin{figure}
%\resizebox{0.75\textwidth}{!}{%
\resizebox{0.5\textwidth}{!}{%
\includegraphics{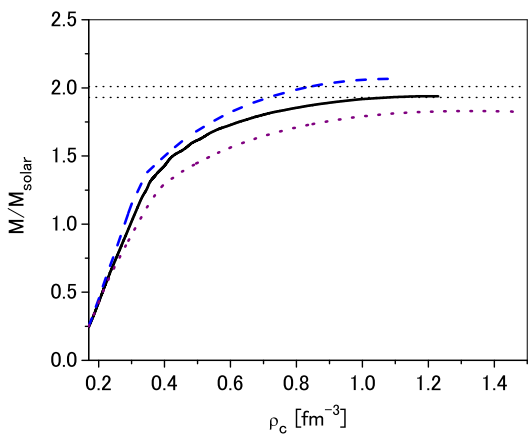}
}
\caption{Neutron-star masses as a function of 
the central density $\rho_c$.
Solid, dashed and dotted curves are for MPa, MPa$^+$ and MPb. 
Two dotted lines show the observed 
mass $(1.97\pm0.04)M_{\odot}$ of J1614-2230.
}
\label{Mrho}
\end{figure}

Using the EoS of hyperonic nucleon matter, we solve the 
TOV equation to obtain the mass and radius of neutron stars. 
The EoS's for MPa, MPa$^+$ and MPb are used $\rho > \rho_0$.
Below $\rho_0$ we use the EoS of the crust obtained
in \cite{Baym1,Baym2}. Then, the EoS's for $\rho > \rho_0$
and $\rho < \rho_0$ are connected smoothly.
%%%%%%%%%%%%%%%%%%%%%%%%%%revise%%%%%%%%%%%%%%%%%%%%%%%%%%%%%%%
%%% revise2
%This value of the matching density is chosen rather arbitrarily.
The behavior of mass-radius curve of a neutron star 
in $\rho_0 \sim 2\rho$ region (below hyperon-onset densities)
is considerably affected by the matching density.
The above value of the matching density is chosen so that
the winding behavior of the mass-radius curve in this region 
becomes as small as possible.
Though there is no physical reason for this choice,
the problem in $\rho_0 \sim 2\rho$ region is not related to 
the EoS and the mass-radius relations in higher-density region, 
which are of our concern in this work.
In Fig.~\ref{MRhyp} (Fig.~\ref{Mrho}), neutron-star masses are 
drawn as a function of radius $R$ (central density $\rho_c$), 
where solid, dashed and dotted curves are for MPa, MPa$^+$ and 
MPb, respectively. Calculated values of maximum masses 
for MPa$^+$, MPa and MPb are 2.07$M_{\odot}$, 1.94$M_{\odot}$ 
and 1.83$M_{\odot}$, respectively, being smaller by 
0.61$M_{\odot}$, 0.51$M_{\odot}$ and 0.35$M_{\odot}$,  
than the values without hyperon mixing. Although the universal
repulsion works to raise the maximum mass, the hyperon mixing 
also is enhanced by it so that the maximum mass is reduced.
This means that the universal repulsion cannot raise the
maximum mass without limit.
The maximum mass for MPb is considerably smaller than
the observed value of $\sim 2M_{\odot}$.
On the other hand, those for MPa and MPa$^+$ reach to
this value owing to the four-body repulsive contributions.
If a neutron star with far heavier mass than $2M_{\odot}$
is observed in future, it may be difficult to reproduce
such a heavy mass in the present modeling for hyperon mixing.

In our calculations, the causality conditions at very high density
are violated in the case of using MPa and MPa$^+$, and not in MPb case.
Then, we adopt the approximation where the EoS is replaced by
the causal EoS above this density in the same way as the treatment 
in \cite{Gandolfi12}.
The critical density for MPa and MPa$^+$, sound speeds being over 
the speed of light, is obtained as 1.2  and 1.1  fm$^{-3}$, respectively.
The masses $M/M_\odot$ take the maximum values at $\sim 1.23$ and 
$\sim 1.10$ fm$^{-3}$ in the case of MPa and MPa$^+$, respectively.
The critical density for the violation of causality condition 
is almost the same as the density giving the maximum mass in these cases.
Therefore, the obtained maximum masses are not so dependent on 
the above approximation. 
On the other hand, the causality condition is violated
significantly in the corresponding result without hyperon mixing.

\subsection{$\Xi\!N$ interaction and $\Xi^-$ mixing}
\label{sec:23}

%\subsection{$\Xi\!N$ interaction}
%\label{sec:31}

Experimental information for $\Xi\!N$ interactions can be obtained from 
emulsion events of simultaneous emission of two $\Lambda$ hypernuclei 
(twin $\Lambda$ hypernuclei) from a $\Xi^-$ absorption point. 
The $\Xi^-$ produced by the $(K^-,K^+)$ reaction
is absorbed into a nucleus ($^{12}$C, $^{14}$N or $^{16}$O
in emulsion) from some atomic orbit, and by the following
$\Xi^-p \rightarrow \Lambda\!\Lambda$ process
two $\Lambda$ hypernuclei are produced.
Then, the energy difference between the initial $\Xi^-$ state 
and the final twin $\Lambda$ hypernuclei gives rise to the binding
energy $B_{\Xi^-}$ between $\Xi^-$ and the nucleus.
Two events of twin $\Lambda$ hypernuclei (I)~\cite{Aoki93}
and (II)~\cite{Aoki95} were observed in the KEK E176 experiment. 
These events were interpreted to be reactions of $\Xi^-$
captured from the $2P$ state in $^{12}$C, 
though another possibility cannot be ruled out.  
In \cite{Ehime}, the strength of the $\Xi\!N$ interaction
was fitted according to this interpretation.
Recently the new event (III)~\cite{Nakazawa} has been 
observed in the KEK E373 experiment, as the first clear evidence 
of a $\Xi^-$ bound state. 
This event is uniquely identified as
%\begin{eqnarray}
%&&\Xi^- + ^{14}{\rm\!N} \rightarrow ^{10}_{\ \Lambda}{\rm\!Be}+ 
%^5_\Lambda{\rm\!He}  \ ,
%\end{eqnarray}
%with $B_{\Xi^-}=4.38 \pm 0.25$ MeV, 
$\Xi^- + ^{14}{\rm\!N} \rightarrow ^{10}_{\ \Lambda}{\rm\!Be}+ 
^5_\Lambda{\rm\!He}$ ,
where $^{10}_{\ \Lambda}{\rm\!Be}$ is in an excited state.
Assuming the $\Xi^-$ capture from the $2P$ state in $^{14}$N,
the obtained value of $\Xi^-$ binding energy $B_{\Xi^-}(2P)$ 
agrees nicely to the value predicted in \cite{Ehime}.

Recently, it has been shown in \cite{ESC08c3} that the 
$\Xi^-$ binding energies extracted from the above events can be
reproduced well by the G-matrix interaction derived from ESC08c.
In Table~\ref{tab:5} we show the potential energies $U_\Xi(\rho_0)$
for a zero-momentum $\Xi$ and their partial-wave contributions 
%in $(T, ^{2S+1}L_J)$ states 
for MPa and ESC. 
As well as the cases of $\Lambda$ and $\Sigma$, the result for MPa 
is similar to that for ESC because of cancellations of MPP 
and TBA contributions in normal density region.

\begin{table}
\caption{Values of $U_\Xi$ at normal density and partial wave
contributions for MPa and ESC (in MeV).
Values specified by $P$ give sum of $(S,J)$ contributions.
}
\label{tab:5}
\begin{tabular}{lr|rrr|r}
\hline\noalign{\smallskip}
&& $^1S_0$ & $^3S_1$ &  $P$\ & $U_\Xi$  \\
\noalign{\smallskip}\hline\noalign{\smallskip}
MPa     & $T=0$   &     1.0 & $-$8.1 &    1.5   &      \\
        & $T=1$   &     9.7 & $-$12.0&    1.1   & $-$6.7 \\
\noalign{\smallskip}
ESC     & $T=0$   &     1.1 & $-$8.0 &    0.9   &      \\
        & $T=1$   &    10.7 & $-$10.8& $-$0.7   & $-$6.8 \\
\noalign{\smallskip}\hline
\end{tabular}
\end{table}

Now, we derive $\Xi\!N$ G-matrices in nuclear matter
composed of $n+p+\Xi^-$ with ESC.
$\Xi\!N$ sectors in ESC are of complicated structure.
For instance, $\Xi\!N$ channels in $T=1$ state couples
to $\Lambda\!\Sigma$ and $\Sigma\!\Sigma$ channels
together with tensor coupling.
For simplicity, we introduce an approximation to replace
them by $\Xi\!N$-$\Xi\!N$ single-channel potentials,
which are determined so as to simulate the results by 
coupled-channel G-matrix calculations at normal-density.
%%%%%%%%%%%%%%%%%%%%%revise%%%%%%%%%%%%%%%%%%%%%%%%%%%%%%%
% revise2
This approximation is for the sake of avoiding difficulties
to solve the G-matrix equations with complicated coupling
structures. For such a G-matrix equation, we have
no good convergence for iterations in high-density region.
%to obtain stable solutions in high-density region,
It is our future problem to develop some technique
to perform accurate calculations in such a case.
%though more accurate calculations are needed in future,
However, the effects of $\Xi$ mixing to the stiffness of EoS
and the mass-radius relation of neutron stars may be not so large
in the cases of including MPP contributions which dominate
the energy densities in high-density region.

\begin{figure}
%\resizebox{0.75\textwidth}{!}{%
\resizebox{0.5\textwidth}{!}{%
\includegraphics{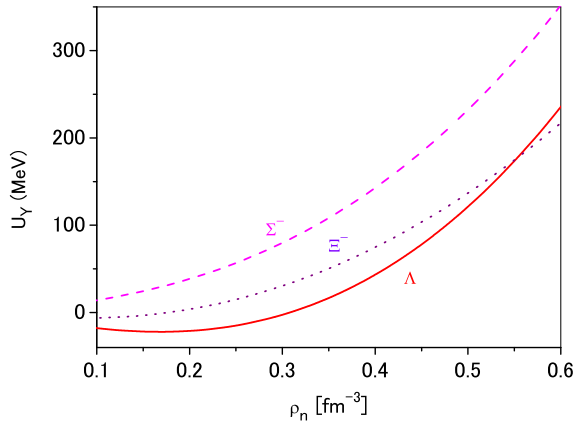}
}
\caption{Single particle potentials $U_Y$ of $Y=\Lambda, \Sigma^-, \Xi^-$ 
for MPa in neutron matter as a function of neutron density $\rho_n$
with $\rho_Y/\rho_n=0.1$.
}
\label{Ehyp}
\end{figure}

Calculations for MPa are performed, including MPP+TBA
parts also in $\Xi\!N$ channels.
In Fig.\ref{Ehyp}, we compare calculated single particle
potentials $U_Y^{(n)}(k=0)$ with $Y=\Lambda, \Sigma^-, \Xi^-$ 
in neutron matter defined by Eq.(\ref{ypot}), where
they are given as a function of neutron density $\rho_n$
in the case of $\rho_Y/\rho_n=0.1$.
The $\Xi^-$ potential turns out to be less repulsive 
than the $\Sigma^-$ potential.

The reason why the $\Sigma^-$ potential is so repulsive is 
because of the Pauli-forbidden effect in the $^3S_1$ $T=3/2$ state.
%istaken into account by strengthening the pomeron coupling~\cite{ESC08,ESC08c2}.
As mentioned before, however, our $\Sigma^-$ potential is less repulsive 
in comparison with the value $U_{\Sigma^-}\sim +30$ MeV assumed in the 
relativistic mean field models~\cite{Weiss,Bednarek,Jiang}. 
This is the reason why the onset density of $\Sigma^-$ is not 
so high in our EoS.

The values of fitted parameters for $\Xi^-$ energy densities
are listed in Table \ref{paraX}.
\begin{table}
\caption{Parameters of $\Xi^-$ energy densities 
given by analytical forms Eq.(13)$\sim$(15) for MPa.}
\label{paraX}
\begin{tabular}{rrrrr}
\hline\noalign{\smallskip}
  $a_{\Xi 0}^{(0)}$ & $a_{\Xi 1}^{(0)}$ &
  $b_{\Xi 0}^{(0)}$ & $b_{\Xi 1}^{(0)}$ & $c_\Xi^{(0)}$ \\
         $-$290.9 &   145.8  & 1112. & 304.2 &  1.893 \\
\noalign{\smallskip}
  $a_{\Xi 0}^{(1)}$ & $a_{\Xi 1}^{(1)}$ &
  $b_{\Xi 0}^{(1)}$ & $b_{\Xi 1}^{(1)}$ & $c_\Xi^{(1)}$ \\
         $-$133.4 &    8.011 & 783.2 & 739.6 &  2.077 \\
\noalign{\smallskip}\hline
\end{tabular}
\end{table}

\begin{figure}
%\resizebox{0.75\textwidth}{!}{%
\resizebox{0.5\textwidth}{!}{%
\includegraphics{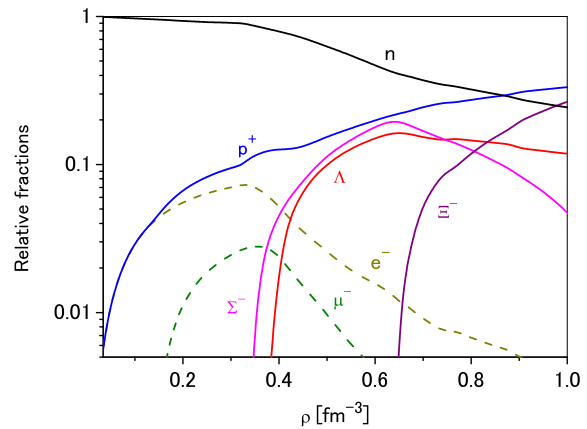}
}
\caption{Composition of hyperonic neutron-star matter 
for MPa.}
\label{chemX}
\end{figure}

In Fig.~\ref{chemX}, the matter compositions are shown 
in the case of MPa. Here, the onset densities of
$\Sigma^-$, $\Lambda$ and $\Xi^-$ are 0.32 fm$^{-3}$,
0.36 fm$^{-3}$ and  0.62 fm$^{-3}$, respectively.

\begin{figure}
%\resizebox{0.75\textwidth}{!}{%
\resizebox{0.5\textwidth}{!}{%
\includegraphics{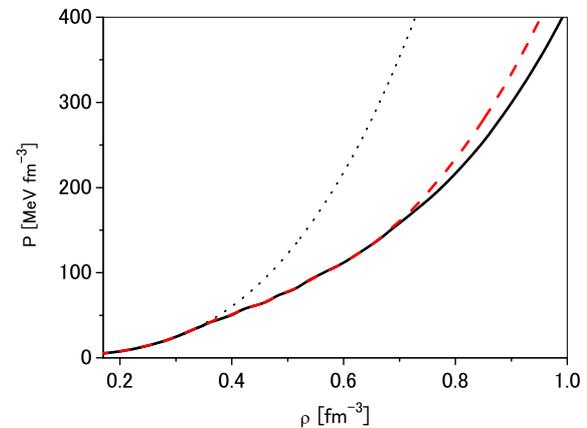}
}
\caption{Pressure $P$ as a function of baryon density $\rho$.
Dashed (solid) curves are with (without) $\Xi^-$ mixing for MPa.
Dotted curve is without hyperon mixing.
}
\label{pressX}
\end{figure}

In Fig.~\ref{pressX}, the calculated values of 
pressure $P$ are drawn as a function of 
baryon density $\rho$ for MPa.
Dashed and Solid curves are with and 
without $\Xi^-$ mixing, respectively. 
Dotted curve is without hyperon mixing.

\begin{figure}
%\resizebox{0.75\textwidth}{!}{%
\resizebox{0.5\textwidth}{!}{%
\includegraphics{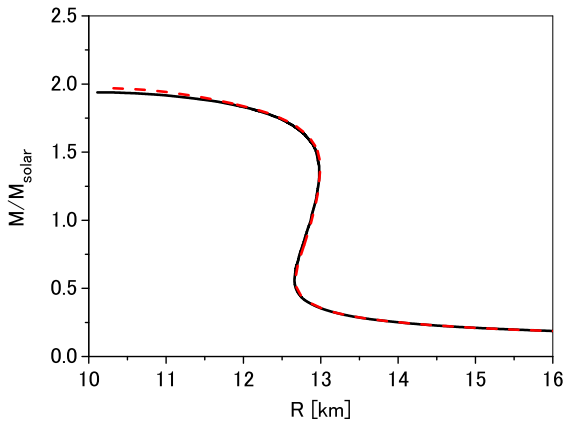}
}
\caption{Neutron-star masses as a function of the radius $R$
derived from the hyperon-mixed EoS for MPa. 
Dashed (solid) curve is with (without) $\Xi^-$ mixing.}
\label{MRhypX}
\end{figure}

In Fig.~\ref{MRhypX}, neutron-star masses are drawn as 
a function of radius, where the dashed (solid) curve is 
obtained from the EoS with (without) the $\Xi^-$ mixing.
The values of maximum masses are 1.97$M_{\odot}$ and
1.94$M_{\odot}$ in the cases of dashed and solid curves.
Thus, the maximum mass is hardly changed, though $\Sigma^-$ 
and $\Lambda$ are replaced by $\Xi^-$ gradually with increase of
baryon density as seen in Fig.~\ref{chemX}. 
The reason is because the universal repulsions (MPP)
work equally to $\Lambda$, $\Sigma^-$ and $\Xi^-$,
being dominant contributions in high density region.
%In Fig.~\ref{Mrhoc}, neutron-star masses are drawn as 
%a function of central density $\rho_c$, 
%where the dashed (solid) curve is obtained from the EoS 
%with (without) the $\Xi^-$ mixing, and the dotted curve
%is without hyperon mixing.
The effect of $\Xi^-$ mixing in Fig.~\ref{MRhypX} can be understood 
correspondingly by the difference between the EoS's with and without 
$\Xi^-$ mixing in Fig.~\ref{pressX}, where the former becomes harder 
than the latter in higher density region. 
%The resultant maximum masses with 
%$\sim 2M_{\odot}$ are very similar to each other. 

%\begin{figure}
%%\resizebox{0.75\textwidth}{!}{%
%\resizebox{0.5\textwidth}{!}{%
%\includegraphics{Mrhoc.eps}
%}
%\caption{Neutron-star masses as a function of the central 
%density $\rho_c$ derived from the hyperon-mixed EoS for MPa. 
%Dashed (solid) curve is with (without) $\Xi^-$ mixing.
%Dotted curve is without hyperon mixing.
%}
%\label{Mrhoc}
%\end{figure}

\section{Conclusion}
\label{sec:3}

The existence of neutron stars with $2M_{\odot}$ gives a severe 
condition for the stiffness of EoS of neutron-star matter.
Though the strong TNR can make the EoS stiff enough,
the hyperon mixing in neutron-star matter brings about the 
remarkable softening of the EoS to cancel the TNR effect.
As a possibility to avoid this serious problem, we introduce
the universal repulsions working for $Y\!N\!N$, $Y\!Y\!N$ 
$Y\!Y\!Y$ as well as for $N\!N\!N$ \cite{NYT}.

On the basis of the $BB$ interaction model ESC, we introduce the
universal three- and four-body repulsion MPP among baryons
together with the phenomenological three-body attraction TBA.
The strengths of MPP in nucleon channels are determined so as 
to fit the observed angular distribution of $^{16}$O+$^{16}$O 
elastic scattering at $E_{in}/A=70$ MeV with use of the G-matrix 
folding potential. The TBA parts are taken so as to assure the 
nuclear saturation property. Then, the stiff EoS of neutron-star
matter is derived on the basis of terrestrial experiments, 
giving the large neutron-star mass over $2M_{\odot}$,
when the hyperon mixing is not taken into account.

In order to study the effect of hyperon mixing to the EoS, 
it is necessary to use reliable interactions in channels 
including hyperons. The reliability of ESC in these channels 
have been confirmed by successful applications to analyses 
of hypernuclear phenomena. The MPP contributions are defined
to exist universally in every baryonic system.
Taking the remaining part TBA also universally,
single particle potentials of $\Lambda$, $\Sigma^-$ and $\Xi^-$
are reproduced consistently with experimental data by using
ESC+MPP+TBA interactions.

The EoS of hyperonic nuclear matter is obtained from ESC+MPP+TBA
on the basis of the G-matrix approach, and the mass-radius relations
of neutron stars are derived by solving the TOV equation.
Though hyperon mixing leads to remarkable softening of the EoS,
the stiffness is partially recovered owing to the MPP contribution.
Quantitatively, in the case of MPb including only the three-body 
repulsion, the derived maximum mass is considerably smaller than
$2M_{\odot}$. In order to reproduce $2M_{\odot}$,
the decisive roles are played by the four-body repulsions
included in MPa and MPa$^+$. 
The effective two-body interactions derived from
the three- and four-body repulsions are proportional to
$\rho$ and $\rho^2$, respectively. 
%The latter contribution in high density region is important 
%to stiffen the EoS enough to give $2M_{\odot}$.
The latter contribution in high density region is sufficient to 
stiffen the EoS enough to give $2M_{\odot}$.
%%%%%%revise%%%%%%%%%%%%%%%%%%%%%%%%%%%%%%%%%%%%%%%%%
However, as the MPP repulsion becomes strong, the resultant
softening effect of EoS becomes large. Namely, the neutron-star
mass does not become far bigger than $2M_{\odot}$ with increase 
of the MPP repulsion: We can say that our three- and four-pomeron 
exchange model provides an upper limit $\sim 2M_{\odot}$ to 
the maximum mass of neutron stars with hyperon mixing.
%(If neutron stars with masses significantly larger than $2M_{\odot}$
%exist, we have to go beyond the quartic multi-pomeron vertices.)\\

The universality of the MPP repulsion not only applies to nucleons 
and hyperons but also baryon resonances ($\Delta_{33}$ etc.) 
and mesons as well. Therefore, the MPP repulsion prevents in general 
the softening of the EoS in the high-density region.

It is interesting to study the effect of $\Xi^-$ mixing
in addition to mixing of $\Lambda$ and $\Sigma^-$
on the basis of the reliable $\Xi\!N$ interaction model.
The emulsion data of twin $\Lambda$ hypernuclei indicate 
that the $\Xi\!N$ interaction is substantially attractive.
Then, the $\Xi^-$ binding energies are reproduced well by ESC.
The EoS of neutron-star matter including hyperons
($\Lambda$, $\Sigma^-$, $\Xi^-$) is derived from ESC+MPP+TBA
interactions. Though there appears considerable $\Xi^-$
mixing in high density region, there is almost no effect 
on the maximum mass.

It should be noted that our conclusion for neutron stars is
obtained essentially on the basis of terrestrial experiments
for nuclear and hypernuclear systems
without using ad hoc parameters to stiffen the EoS.
It can be said, at least, that our approach contributes to 
one of possible solutions of the hyperon puzzle.

%\section{Section title}
%\label{sec:1}
%and \cite{RefJ}
%\subsection{Subsection title}
%\label{sec:2}
%as required. Don't forget to give each section
%and subsection a unique label (see Sect.~\ref{sec:1}).
%\begin{figure}
%\resizebox{0.75\textwidth}{!}{%
%  \includegraphics{leer.eps}
%\vspace{5cm}       % Give the correct figure height in cm
%\end{figure}
%\begin{figure*}
%\vspace*{5cm}       % Give the correct figure height in cm
%\end{figure*}
%\begin{table}
%\begin{tabular}{lll}
%\hline\noalign{\smallskip}
%\end{tabular}
% Or use
%\vspace*{5cm}  % with the correct table height
%\end{table}
%
% BibTeX users please use
% \bibliographystyle{}
% \bibliography{}

\begin{thebibliography}{}

\bibitem{Demorest10}
P.B. Demorest, T. Pennucci, S.M. Ransom, M.S.E. Roberts, and J.W. Hessels,
Nature (London) {\bf 467}, (2010) 1081. 

\bibitem{Antoniadis13}
J. Antoniadis {\it et al.},
Science {\bf 340}, (2013) 6131.

\bibitem{Baldo00}
M. Baldo, G.F. Burgio, and H.-J. Schulze,
Phys. Rev. C{\bf 61}, (2000) 055801. 

\bibitem{Vidana00}
I. Vidana, A. Polls, A. Ramos, L. Engvik, and M. Hjorth-Jensen,
Phys. Rev. C{\bf 62}, (2000) 035801. 

\bibitem{NYT}
S. Nishizaki, Y. Yamamoto, and T. Takatsuka,
Prog. Theor. Phys. {\bf105}, (2001) 607; {\bf 108}, (2002) 703.

\bibitem{YFYR13}
Y.Yamamoto, T.Furumoto, N.Yasutake and Th.A.Rijken,
Phys. Rev. C{\bf 88}, (2013) 022801. 

\bibitem{YFYR14}
Y.Yamamoto, T.Furumoto, N.Yasutake and Th.A.Rijken,
Phys. Rev. C{\bf 90}, (2014) 045805. 

\bibitem{ESC08} 
Th.A. Rijken, M.M. Nagels, and Y. Yamamoto, 
Prog. Theor. Phys. Suppl. {\bf 185}, (2010) 14. 

\bibitem{ESC08c1} 
M.M. Nagels, Th.A. Rijken, and Y. Yamamoto, 
arXiv:1408.4825 (2014).

\bibitem{ESC08c2} 
M.M. Nagels, Th.A. Rijken, and Y. Yamamoto, 
arXiv:1501.06636 (2015).

\bibitem{ESC08c3} 
M.M. Nagels, Th.A. Rijken, and Y. Yamamoto, 
arXiv:1504.02634 (2015).

\bibitem{FSY}
T. Furumoto, Y. Sakuragi, and Y. Yamamoto, 
Phys. Rev. C{\bf 79}, (2009) 011601(R); C{\bf 80}, (2009) 044614.

\bibitem{FSY14} 
T. Furumoto, Y. Sakuragi, and Y. Yamamoto, 
Phys. Rev. C{\bf 90}, (2014) 041601(R).

\bibitem{Panda81}
I.E. Lagaris and V.R. Pandharipande,
Nucl. Phys. A{\bf 359}, (1981) 349.

\bibitem{Nakazawa} K. Nakazawa {\it et al.}, 
Prog. Theor. Exp. Phys. {\bf 2015}, 033D02.

\bibitem{Baldo98}
H.Q.Song, M.Baldo, G.Giansiracusa and U.Lombardo,
Phys. Rev. Lett. {\bf 81}, (1998) 1584. 

\bibitem{Baldo02}
M.Baldo, A.Fiasconaro, H.Q.Song, G.Giansiracusa and U.Lombardo,
Phys. Rev. {\bf C65}, (2002) 017303. 

%\bibitem{SJM} 
%R. Tamagaki,
%Prog. Theor. Phys. {\bf 119} (2008) 905.

\bibitem{Kai74}
A.B. Kaidalov and K.A. Ter-Materosyan, 
Nucl. Phys. {\bf 74}, (1974) 471.

\bibitem{Bron77}
J.B. Bronzan and R.L. Sugar, 
Phys. Revs. D{\bf 16}, (1977) 466.

\bibitem{Nuoffer}
F. Nuoffer, et al., 
Nuovo Cimento A{\bf 111}, (1998) 971.


\bibitem{UIX}
B.S. Pudliner, V.R. Pandharipande, J. Carlson, and R.B. Wiringa,
Phys. Rev. Lett. {\bf 74}, (1995) 59. 

\bibitem{Skyrme}
Y. Yamamoto, H. Band\={o} and J. \u{Z}ofka,
Prog. Theor. Phys. {\bf 80}, (1988) 757 ;
D.J. Millener, C.B. Dover and A. Gal,
Phys. Rev. C{\bf 38}, (1988) 2700.  

\bibitem{Schlze00}
J. Cugnon, A. Lejeune and H.-J. Schulze,
Phys. Rev. C{\bf 62}, (2000) 064308.

\bibitem{Harada05}
T. Harada and Y. Hirabayashi,
Nucl. Phys. A{\bf 759}, (2005) 143.

\bibitem{Baym1}
G. Baym, A. Bethe, and C. Pethick,
Nucl. Phys. A{\bf 175}, (1971) 225.

\bibitem{Baym2}
G. Baym, C.J. Pethick, and P. Sutherland,
Astrophys. J.{\bf 170}, (1971) 299.

\bibitem{Gandolfi12}
S. Gandolfi, J. Carlson, and Sanjjay Reddy,
Phys. Rev. C{\bf 85}, (2012) 032801(R). 


\bibitem{Aoki93} S. Aoki {\it et al.}, 
Prog. Theor. Phys. {\bf 89}, (1993), 493.

\bibitem{Aoki95} S. Aoki {\it et al.}, 
Phys. Lett. {\bf B355}, (1995), 45.

\bibitem{Ehime} 
M. Yamaguchi, K. Tominaga, Y. Yamamoto, and T. Ueda, 
Prog. Theor. Phys. {\bf 105}, (2001), 627.

\bibitem{Weiss}
S. Weissenborn, D. Chatterjee, and J. Schaffner-Bielich,
Nucl. Phys. A{\bf 881}, (2012) 62.

\bibitem{Bednarek}
I. Bednarek, P. Haensel, J.L. Zdunik, M. Bejger, and R. Ma\'{n}ka,
Astronomy \& Astrophysics A{\bf 157}, (2012) 543.

\bibitem{Jiang}
Wei-Zhou Jiang, Bao-An Li, and Lie-Wen Chen,
Astrophys. J.{\bf 756}, (2012) 56.


%Author, \textit{Book title} (Publisher, place year) page numbers
% etc
\end{thebibliography}

\end{document}